\documentstyle[twoside,fleqn,epsfig,espcrc2]{article}
\newcommand{\ee}{\end{enumerate}}        
\newcommand{\bi}{\begin{itemize}}
\newcommand{\ei}{\end{itemize}}        
\newcommand{\eq}[1]{{\protect\frenchspacing Eq~(\ref{#1})}}

\newcommand{\beq}{\protect\begin{equation}}
\newcommand{\eeq}{\protect\end{equation}}        
\newcommand{\bqa}{\begin{eqnarray}}        
\newcommand{\eqa}{\end{eqnarray}}        
\newcommand{\be}{\begin{enumerate}}

\title{Compact QED under scrutiny: it's first order} 
\author{G.~Arnold$^{1}$, B.~Bunk$^{2}$, Th.~Lippert$^{1}$,
K.~Schilling$^{1}$ \\ {\small $^{1}$Department of Physics, University
of Wuppertal, D-42097 Wuppertal, Germany}\\ {\small$^{2}$Department of
Physics, Humboldt University Berlin, D-10099 Berlin, Germany}}
\begin{document}
\begin{abstract}
We report new results from our finite size scaling analysis of 4d
compact pure U(1) gauge theory with Wilson action. Investigating
several cumulants of the plaquette energy within the Borgs-Kotecky
finite size scaling scheme we find strong evidence for a first-order
phase transition and present a high precision value for the critical
coupling, $\beta_T$ in the thermodynamic limit.
\end{abstract}
\maketitle

\section{INTRODUCTION}
The nature of the phase transition in compact QED has been under debate for
long time. We have addressed this problem in high statistics runs to
reach a final conclusion in that matter. An important ingredient of
our approach is the finite size scaling (FSS) theory \`a la
Borgs-Kotecky (BK) first established a long time ago in the context of
strong first order phase transitions
\cite{BORGS1,BORGS2,BORGS3}. According to BK the finite volume
partition function at temperature $\beta$ in finite volumes with
periodic boundary conditions (neglecting interfacial contributions)
has the remarkably simple form $
Z=e^{-Vf(\beta)}=e^{-Vf_1(\beta)}+e^{-Vf_2(\beta)+\ln(X)}$.  The
functions $f_1(\beta)$ and $f_2(\beta)$ denote bulk free energy
densities in the two coexisting phases 1 and 2.  $X$ stands for the
asymmetry parameter which is nothing but the relative phase weight in
the probability distribution $P(E)$.

A heuristic extension of the BK ansatz to weak first order transition
was demonstrated for the 3d 3-state Potts model \cite{JANKE97}. 
The conclusion of our work is based on a validation of BK by
perturbative analysis as well as independent ab initio determinations
of the gap characteristics.
\section{SIMULATION DETAILS}
We consider 4d pure U(1) gauge theory with Wilson action $S =
-\beta\sum_{n,\nu>\mu}\cos(\theta_{\mu\nu}(n))$, where $\beta$
represents the Wilson coupling and $\theta_{\mu\nu}(n)$ the plaquette
angle. We use a lattice of volume $V=L^4$ with periodic boundary
conditions.

We have implemented three different algorithms for generating the U(1)
gauge field configurations: (a) a local Metropolis (Metro), updating
each link separately, (b) a global hybrid Monte Carlo algorithm (HMC)
and (c) a combination of the multicanonical and the hybrid Monte Carlo
algorithm (MHMC). For details we refer to Refs.
\cite{ARNOLD98,ARNOLD99}. The cumulative number of generated
configurations at each lattice size $L$ is $>5 \times 10^6$. We
measure the number of tunneling events (flips) as control parameter
for the mobility of the algorithms and the integrated plaquette
autocorrelation time $\tau_{int}$ which controls the statistical
quality of each single Markov chain. Runs differing by coupling,
algorithm, HMC parameters or by weight function are considered as
independent. Our simulation parameters are listed in Table
\ref{tab:mcit}.

\begin{table}[htb]
\caption{Simulation details. The HMC subscript denotes length of trajectory.}
\center
\tiny
\begin{tabular}{|r|l|c|r|r|r|}
\hline
 L  & $\beta$ & algorithm & \#conf$\times 10^6$ & \#flips &  $\tau_{int}$ \\
\hline
\hline

6& 1.001700&    Metro&  11.20&  21170          &104(2) \\
& 1.001500&    Metro&  9.80&   18300           &104(2) \\
& 1.001600&     HMC&    2.48& 3577             &130(5) \\
& 1.001772&    MHMC&    4.90& 7444             &102(2) \\
\hline

8& 1.007370&    Metro&   2.79&2512             &304(16)\\
&   1.007370&   HMC$_2$&     1.25&274         &1256(196)\\
&   1.007370&   HMC$_4$&     1.25&554          &649(73)\\
&   1.007370&   HMC$_6$&     1.25&698          &450(42)\\
&   1.007370&   HMC$_8$&     1.25&846          &390(33)\\
&   1.007370&   HMC$_9$&     1.25&907          &339(27)\\
&   1.007370&   HMC$_{10}$&     1.25&921       &328(26)\\
&   1.007370&   HMC$_{11}$&     1.25&929       &345(28)\\
&   1.007370&   HMC$_{12}$&     1.25&911       &363(30)\\
&   1.007370&   HMC$_{13}$&     1.25&932       &355(29)\\
&   1.007370&   HMC$_{14}$&     1.25&854       &382(33)\\
&   1.007370&   HMC$_{16}$&     1.25&856       &399(35)\\
&   1.007370&   HMC&     1.44&  1058           &379(30)\\
&   1.007337&    MHMC&         6.36&5179       & 240(7)\\
\hline

10& 1.009300&   Metro&  4.37&   1770           &784(52)\\
  & 1.009400&   Metro&  7.44&   3104           &775(35)\\
  & 1.009300&   HMC$_9$&  1.00&   294          &948(144)\\
  & 1.009300&   HMC$_{11}$&  1.00&  350        &897(132)\\
  & 1.009300&   HMC$_{15}$&  1.00&  353        &1060(170)\\
  & 1.009300&   HMC$_{17}$&  1.00&  344        &831(118)\\
  & 1.009300&   HMC$_{19}$&  1.00&  340        &894(132)\\
 &  1.009300&    MHMC &   2.61& 1318           & 412(56)\\
\hline

12& 1.010143&  Metro&    3.62&569              &2406(304)\\
 &  1.010143&  Metro&    5.88&916              &2058(189)\\
 &  1.010143&  Metro&    1.75&315              &2576(486)\\
 &  1.010143&  Metro&    1.87&308              &2098(345)\\
 &  1.010143&   MHMC&         1.30&403         & 734(86)\\
 &  1.010143&   MHMC&         2.18&702         & 689(60)\\
\hline

14& 1.010598&  Metro&    3.90& 215             &5980(1150)\\
&   1.010600&  Metro&    6.52&395              &7480(1240)\\
&   1.010568&  HMC  &    0.57&26               &12500(9000)\\
&   1.010568&   MHMC  &   0.83&169             & 1070(190)\\
&   1.010568&   MHMC  &   3.80&725             & 1380(130)\\
\hline

16& 1.010753&  Metro&    5.42&75               &22900(9000)\\
&   1.010800&  Metro&    5.55&93               &25400(6800)\\
&   1.010753&   MHMC &    0.60&66              & 1980(560)\\
&   1.010753&   MHMC &    0.63&77              & 1850(500)\\
&   1.010753&   MHMC &    1.63&151             & 1770(290)\\
&   1.010753&   MHMC &    3.39&424             & 1800(200)\\
\hline

18&   1.010900&  MHMC &    0.38&13             & 9900(7900)\\
&   1.010900&   MHMC &    0.63&20              & 10000(6800)\\
&   1.010900&   MHMC &    0.98&58              & 5800(2200)\\
&   1.010900&   MHMC &    3.79&272             & 4700(820)\\
\hline
\end{tabular}
\label{tab:mcit}
\end{table}

\section{CUMULANTS}
Based on the plaquette operator 
\beq E=\frac{1}{6V}\sum_{n,\nu>\mu}\cos(\theta_{\mu\nu}(n)), \eeq
we consider the following cumulants:
\begin{eqnarray*}
  C_{v}(\beta,L) & =& %
  6V\left(\left<E^{2}\right> - \left<E\right>^{2}\right), \label{cumulant1}\\
  U_{2}(\beta,L) & = &%
  1 - \frac{\left<E^{2}\right>}{\left<E\right>^{2}},\\
  U_{4}(\beta,L) & = &%
  \frac{1}{3}\left(1-\frac{\left<E^{4}\right>}{\left<E^{2}\right>^{2}}\right).
\end{eqnarray*}
In addition to their derivatives with respect to $\beta$ we measure
higher derivatives of the free energy density $(-1)^{n+1}
\kappa_{n}(\beta,L) =
\frac{\partial^{n}f(\beta,L)}{\partial\beta^{n}}$. Introducing the
central moments $\mu_{n} =
V^{n-1}\left<\left(E-\left<E\right>\right)^{n}\right>$
we can write them as 
\begin{eqnarray*}
\kappa_{3} & = & \mu_{3},\\
   \kappa_{4} & = & \mu_{4} -3V\mu_{2}^{2}, \\
   \kappa_{5} & = & \mu_{5} -10V\mu_{2}\mu_{3}, \\
   \kappa_{6} & = & \mu_{6} -15V\mu_{2}\mu_{4}%
   -10V\mu_{3}^{2} +30V^{2}\mu_{2}^{3}.
\end{eqnarray*}

For each of the ten cumulants the location $(\beta_\kappa, \kappa)$ of
its rightmost extremum is determined by reweighting the measured
probability distribution $P(E)$ to different couplings $\beta$. To
calculate the estimates of our cumulants at each lattice size $L$ we
proceed in two steps: i) we determine the error of each individual run
performing a jackknife error analysis by subdivision of the run into
ten blocks; ii) we calculate the final result by $\chi^2$-fitting
these individual results to a constant.

\section{FINITE SIZE SCALING}
Let us consider the expansion of the pseudocritical coupling for
$C_v$; in the BK representation it is given by
\beq \beta_{Cv}(V)=\beta_{Cv}(\infty)+\sum_{k=1}^{k_{max}} B_kV^{-k}.
\label{eqbeta} \eeq
In order to expose systematic effects in the fit parameter
$\beta_{Cv}(\infty)$, we vary both, the fit range within $L_{min}\le L
\le 18$ and the truncation parameter $k_{max}$. Table \ref{tabfitcmax}
displays a remarkable stability pattern both for $\beta_{Cv}$ and
$B_1$ supporting the validity of the $V^{-1}$-expansion. Averaging
the best fit couplings to a constant we obtain $\beta_{Cv}(\infty)=
1.0111310(62)$. Performing the same analysis for all ten cumulants
yields an average infinite volume transition coupling \beq
\beta_T=1.0111331(21) \label{beta_T}. \eeq
\begin{table}[hbt]
\caption{Transition couplings $\beta_{Cv}(\infty)$ fitted to
\eq{eqbeta}. Best fits are in bold face letters.}
\center
\small
\begin{tabular}{|r|l|l|l|l|}
\hline
 $L_{min}$  &  $k_{max}$ & $\chi^2_{dof}$ &
$\beta_{Cv}(\infty)$ & $B_1$\\
\hline
\hline
14 &    1& 1.03& 1.0111241(13) & -18.95(14)\\
\hline
12 &    1&1.09& 1.0111144(55) & -18.24(21)\\
   &    2&\bf  0.19&\bf 1.0111315(57)&\bf -19.96(53)\\
\hline
10 &     1& 12.7& 1.0110945(147)& -17.18(37)\\
   &     2&\bf 0.13&\bf 1.0111283(25)&\bf -19.63(15)\\
   &     3&\bf 0.21&\bf 1.0111319(62)&\bf -20.06(65)\\
\hline
8  &     1&108& 1.0110474(349)& -15.33(50)\\
   &     2&2.14&1.0111159(69) & -18.70(25)\\
   &     3&\bf 0.11&\bf 1.0111309(25) &\bf -19.94(17)\\
   &     4&\bf 0.21&\bf 1.0111309(22) &\bf -19.94(10)\\
\hline
6  &     1& 970 & 1.0109389(913)& -12.38(56)\\
   &     2& 37.1& 1.0110792(218)& -16.84(41)\\
   &     3&1.25& 1.0111199(55) & -19.02(22)\\
   &     4&\bf 0.10&\bf 1.0111316(11) &\bf -20.02(6)\\
\hline
\end{tabular}
\label{tabfitcmax}
\end{table}

Analogously the expected scaling of the maxima of the specific heat
yields a prediction for the infinite volume gap $G$ \bqa
\frac{C_{v,max}(V)}{6V}&=&\frac{1}{4}G^2+\sum_{k=1}^\infty C_kV^{-k}
\nonumber\\ G&=& 0.026721(59). \label{G} \eqa Scaling of the Binder cumulant yields
\bqa U_{4,min}&=&U+\sum_{k=1}^\infty A_kV^{-k} \nonumber \\ U&=&-5.816(27)\
10^{-4}. \label{U} \eqa From $B_1=-\frac{\ln(X)}{6G}$ we can derive an asymmetry $
\ln(X)=3.21(10). \label{asym1}$

\section{CONSISTENCY CHECKS}
An independent leading order perturbative lattice calculation confirms
the value of $\ln(X)$ without relying on the validity of the BK ansatz
\cite{ARNOLD02}. In the Coulomb phase the partition function can
approximately be written as \[ Z \simeq
\left(\frac{2\pi}{e_R^2}\right)^{-\frac{3}{2}(V-1)} V^\frac{1}{2}
{\prod_p}' \left[ \sum_\mu 2(1-\cos p_\mu) \right]^{-1}, \] leading to
the free energy $F_2=\ln
Z=-\frac{3}{2}(V-1)\\\ln(\frac{2\pi}{e_R^2})+2\ln L-{\sum_p}'\ln
\sum_\mu2(1-\cos p_\mu)$.  The summation over all momenta $p_\mu$ can
be done for asymptotically large $L$ as $ {\sum_p}' \ln \sum_\mu
2(1-\cos p_\mu) = aV + 2 \ln L - b +{\cal O}(L^{-2})$ with parameters
$a = 1.999708$ and $b = 1.701216$ that can be numerically computed to
arbitrary precision. With $ F_{2}(\beta,L) = V f_{2}(\beta) + \Delta
F_2(\beta,L)$ we obtain for the Coulomb phase finite size correction
$\Delta F_2 =b + \frac{3}{2}\ln(2\pi/e_R^2)=3.15(8).$ $e_R$ is taken
from a very accurate measurement of the renormalized fine structure
constant at the phase transition $\alpha_T=e_R^2/4\pi=0.19(1)$
\cite{Jersak97}. Note that the logarithmic correction cancels out for
the 4d symmetric system with periodic boundary conditions.

One can argue that the leading finite size effects in the confined
phase due to $0^{++}$ gaugeballs and string states are at least 3
respectively 6 orders of magnitude smaller than $\Delta F_2$ on a
lattice as small as $L=16$ \cite{ARNOLD02}.  Thus we neglect these
contributions and obtain, in perfect agreement, an asymmetry parameter
$ \ln(\hat{X})= \Delta F_2=3.15(8)$.

Furthermore our very accurate value of $\beta_T$ (\eq{beta_T}) admits
a direct measurement of latent heat and Binder cumulant, performing
metastable simulations at $\beta_T$ on lattice sizes up to $32^4$
\cite{ARNOLD00}. Denoting the energy peaks of the probability
distribution in the confined and Coulomb phases by $E_1(L)$ and
$E_2(L)$ we fit their continuum limit values $E_i=E_i(\infty)$ and
find \bqa \hat{G}&=&E_2-E_1=0.026685(54),\\ \hat{U}&=&
-\frac{(E_2^2-E_1^2)^2}{12E^2_2E^2_1}=-5.777(16)\ 10^{-4}.  \eqa Both
values are in perfect agreement with the BK results from
Eqs~(\ref{G},\ref{U}).
\section{SUMMARY AND CONCLUSION}
All cumulants investigated in our high statistics analysis at
L=6,8,10,12,14,16,18 can be described by BK first-order FSS. Ab initio
measurements {\it confirm} the FSS results for the infinite volume gap
$G$, the Binder cumulant $U$ and the asymmetry $\ln(X)$. The non
vanishing values for $G$ and $U$ lead to the conclusion that the phase
transition in compact 4d $U(1)$ theory with Wilson action is
first-order.

\end{document}